\documentclass[runningheads]{llncs}
\usepackage{xcolor}

\usepackage{graphicx}

\usepackage{amsmath,amssymb}
\graphicspath{ {./images/} }

\begin{document}

\title{Motion Compensated Unsupervised Deep Learning for 5D MRI\thanks{ This work is supported by grants NIH R01 AG067078 and R01 EB031169.}}

%
%\titlerunning{Motion Compensated 5D MRI}

\author{Joseph Kettelkamp\inst{1}\orcidID{0000-0002-9702-5911} \and
Ludovica Romanin\inst{2}\orcidID{0000-0001-5031-3302}\and
Davide Piccini\inst{2}\orcidID{0000-0003-4663-3244} \and
Sarv Priya\inst{1}\orcidID{0000-0003-2442-1902} \and
Mathews Jacob\inst{1}\orcidID{0000-0001-6196-3933}}

%index{Kettelkamp, Joseph}
%index{Romanin, Ludovica}
%index{Piccini, Davide}
%index{Priya, Sarv}
%index{Jacob, Mathews}

%

\authorrunning{J. Kettelkamp et al.} 


\institute{University of Iowa, IA \\\email{\{joseph-kettelkamp, sarv-priya,mathews-jacob\}@uiowa.edu}\and
Advanced Clinical Imaging Technology, Siemens Healthineers International AG, Lausanne, Switzerland\\
\email{\{ludovica.romanin, davide.piccini\}@siemens-healthineers.com}
}

%
\maketitle              % typeset the header of the contribution
\begin{abstract}

We propose an unsupervised deep learning algorithm for the motion-compensated reconstruction of 5D cardiac MRI data from 3D radial acquisitions.  Ungated free-breathing 5D MRI simplifies the scan planning, improves patient comfort, and offers several clinical benefits over breath-held 2D exams, including isotropic spatial resolution and the ability to reslice the data to arbitrary views. However, the current reconstruction algorithms for 5D MRI take very long computational time, and their outcome is greatly dependent on the uniformity of the binning of the acquired data into different physiological phases. The proposed algorithm is a more data-efficient alternative to current motion-resolved reconstructions.  This motion-compensated approach models the data in each cardiac/respiratory bin as Fourier samples of the deformed version of a 3D image template.  The deformation maps are modeled by a convolutional neural network driven by the physiological phase information. The deformation maps and the template are then jointly estimated from the measured data. The cardiac and respiratory phases are estimated from 1D navigators using an auto-encoder.  The proposed algorithm is validated on 5D bSSFP datasets acquired from two subjects.

\end{abstract}

\section{Introduction}

Magnetic Resonance Imaging (MRI) is currently the gold standard for assessing cardiac function.  It provides detailed images of the heart's anatomy and enables accurate measurements of parameters such as ventricular volumes, ejection fraction, and myocardial mass.  Current clinical protocols, which rely on serial breath-held imaging of the different cardiac slices with different views, often require long scan times and are associated with reduced patient comfort.  Compressed sensing \cite{Bustin2020}
, deep learning \cite{Schlemper2018,Oscanoa2023}
, and motion-compensated approaches \cite{Usman2013}
were introduced to reduce the breath-hold duration in cardiac CINE MRI. Unfortunately, many subject groups, including pediatric and older subjects, cannot comply with even the shorter breath-hold durations. 

5D free-breathing MRI approaches that rely on 3D radial readouts \cite{Feng2017,DiSopra2019} have been introduced to overcome the above challenges. These methods resolve the respiratory and cardiac motion from either the center of k-space or Superior-Inferior (SI) k-space navigators.  The k-space data is then binned into different cardiac/respiratory phases and jointly reconstructed using compressed sensing.  The main benefit of this motion-resolved strategy is the ability to acquire the whole heart with isotropic spatial resolution as high as 1mm$^3$.  This approach allows the images to be reformatted into different views to visualize specific anatomical regions at different cardiac and/or respiratory phases.  Despite the great potential of 5D MRI, current methods have some challenges that limit their use in routine clinical applications.  Firstly, the motion-resolved compressed sensing reconstruction is very computationally intensive, and it can take several hours to have a dynamic 3D volume.  And secondly, compressed sensing reconstructions require fine tuning of several regularization parameters, which greatly affect the final image quality, depending on the undersampling factor and the binning uniformity. 

The main focus of this work is to introduce a motion-compensated reconstruction algorithm for 5D MRI.  The proposed approach models the images at every time instance as a deformed version of a static image template.  Such an image model may not be a good approximation in 2D schemes \cite{Usman2013},
where the organs may move in and out of the slice.  However, the proposed model is more accurate for the 3D case.  We introduce an auto-encoder to estimate the cardiac and respiratory phases from the superior-inferior (SI) k-t space navigators.  We disentangle the latent variables to cardiac and respiratory phases by using the prior information of the cardiac and respiratory rates.  The latent variables allow us to bin the data into different cardiac and respiratory phases.  We use an unsupervised deep learning algorithm to recover the image volumes from the clustered data.  The algorithm models the deformation maps as points on a smooth low-dimensional manifold in high dimensions, which is a non-linear function of the low-dimensional latent vectors.  We model the non-linear mapping by a Convolutional Neural Network (CNN).   When fed with the corresponding latent vectors, this CNN outputs the deformation maps corresponding to a specific cardiac or respiratory phase. We learn the parameters of the CNN and the image template from the measured k-t space data. We note that several manifold based approaches that model the images in the time series by a CNN were introduced in the recent years. \cite{Zou2021,Mohsin2019,Rusho2022}. All of these methods rely on motion resolved reconstruction, which is conceptually different from the proposed motion compensated reconstruction.

We validate the proposed scheme on cardiac MRI datasets acquired from two healthy volunteers. The results show that the approach is capable of resolving the cardiac motion, while offering similar image quality for all the different phases.  In particular, the motion-compensated approach can combine the image information from all the motion states to obtain good quality images. 

\section{Methods}

\subsection{Acquisition scheme}

In vivo acquisitions were performed on a 1.5T clinical MRI scanner (MAGNETOM Sola, Siemens Healthcare, Erlangen, Germany). The free-running research sequence used in this work is a bSSFP sequence, in which all chemically shift-selective fat saturation pulses and ramp-up RF excitations were removed, in order to reduce the specific absorption rate (SAR) and to enable a completely uninterrupted acquisition \cite{Roy2022}. K-space data were continuously sampled using a 3D golden angle kooshball phyllotayis trajectory \cite{Piccini2011}, interleaved with the acquisition of a readout oriented along the superior–inferior (SI) direction for cardiac and respiratory self‐gating \cite{DiSopra2019}. The main sequence parameters were: radio frequency excitation angle of 55 with an axial slab-selective sinc pulse, resolution of 1.1 mm3, FOV of 220 mm3, TE/TR of 1.87/3.78 ms, and readout bandwidth of 898 Hz/pixel.
The total fixed scan time was 7:58 minutes. 

%The data was acquired with a research free-running 3D golden-angle radial bSSFP sequence with no cardiac and respiratory gating and no contrast administration.  The parameters of the sequence are TR/TE=4ms/0.3ms. The acquisitions are segmented into multiple interleaves using a spiral phyllotaxis trajectory \cite{Piccini2011}.  All chemically shift-selective fat saturation pulses and ramp-up RF excitations were removed, reducing the specific absorption rate (SAR) and enabling a completely uninterrupted acquisition \cite{Roy2022}.  Each interleaf was preceded by a readout oriented along the superior–inferior (SI) direction for cardiac and respiratory self‐gating \cite{DiSopra2019}. 

\subsection{Forward model}
We model the measured k-space data at the time instant $t$ as the multichannel Fourier measurements of $\boldsymbol\rho_t = \boldsymbol \rho(\mathbf r,t)$, which is the image volume at the time instance $t$:
\begin{equation}
    \mathbf b_t = \underbrace{\mathbf{F}_{\mathbf k_t}\, \mathbf C \boldsymbol \,\boldsymbol\rho_t}_{\mathcal A_t(\boldsymbol\rho_t)}
\end{equation}
Here, $\mathbf C$ denotes the multiplication of the images by the multi-channel coil sensitivities, while $\mathbf F_{\mathbf k}$ denotes the multichannel Fourier operator. $\mathbf k_t$ denotes the k-space trajectory at the time instant $t$.In this work, we group 22 radial spokes corresponding to a temporal resolution of $88$ ms. 

An important challenge associated with the bSSFP acquisition without intermittent fat saturation pulses is the relatively high-fat signal compared to the myocardium and blood pool.  Traditional parallel MRI and coil combination strategies often result in significant streaking artifacts from the fat onto the myocardial regions, especially in the undersampled setting considered in this work. We used the coil combination approach introduced in \cite{Kim2021} to obtain virtual channels that are maximally sensitive to the cardiac region.  A spherical region covering the heart was manually selected as the region of interest (ROI), while its complement multiplied by the distance function to the heart was chosen as the noise mask. We chose the number of virtual coils that preserve 75\% of the energy within the ROI. This approach minimizes the strong fat signals, which are distant from the myocardium. We used the JSENSE algorithm \cite{Ying2007,Uecker2008} to compute the sensitivity maps of the virtual coils.
 
%Ying, L., & Sheng, J. (2007). Joint image reconstruction and sensitivity estimation in SENSE (JSENSE). Magnetic Resonance in Medicine, 57(6), 1196-1202.

%Uecker, M., Hohage, T., Block, K. T., & Frahm, J. (2008). Image reconstruction by regularized nonlinear inversion- joint estimation of coil sensitivities and image content. Magnetic Resonance in Medicine, 60(#), 674-682.

\subsection{Image and motion models}
The overview of the proposed scheme is shown in Fig. 1. The recovery of $\rho_t$ from very few of their measurements $\mathbf b_t$ is ill-posed. To constrain the recovery, we model $\boldsymbol \rho_t$ as the deformed version of a static image template $\boldsymbol \eta(\mathbf r)$:
\begin{equation}\label{def}
    \rho(\mathbf r,t) = \mathcal I \left[\boldsymbol \eta,\phi_t(\mathbf r) \right]
\end{equation}
Here, $\boldsymbol\phi_{t}$ is the deformation map and the operator $\mathcal I$ denotes the deformation of $\boldsymbol \eta$. We implement \eqref{def} using cubic Bspline interpolation. This approach allows us to use the k-space data from all the time points to update the template, once the motion maps $\boldsymbol\phi_t$ are estimated.  

Classical MoCo approaches use image registration to estimate the motion maps $\boldsymbol\phi_t$ from approximate (e.g. low-resolution) reconstructions of the images $\rho(\mathbf r,t)$. However, the quality of motion estimates depends on the quality of the reconstructed images, which are often low when we aim to recover the images at a fine temporal resolution (e.g. 88 ms). 

We propose to estimate the motion maps directly from the measured $k-t$ space data.  In particular, we estimate the motion maps $\boldsymbol\phi_t$ such that the multi-channel measurements of $\rho(\mathbf r,t)$ specified by \eqref{def} match the measurements $\mathbf b_t$. We also estimate the template $\boldsymbol\eta$ from the k-t space data of all the time points.  To constrain the recovery of the deformation maps, we model the deformation maps as the output of a convolutional neural network 
\begin{equation*}
    \phi_t = \mathcal G_{\theta}[\mathbf z_t],
\end{equation*}
in response to low-dimensional latent vectors $\mathbf z_t$.  Here, $\mathcal G_{\theta}$ is a convolutional neural network, parameterized by the weights $\theta$.  We note that this approach constrains the deformation maps as points on a low-dimensional manifold. They are obtained as non-linear mappings of the low-dimensional latent vectors $\mathbf z_t$, which capture the motion attributes. The non-linear mapping itself is modeled by the CNN.

\subsection{Estimation of latent vectors from SI navigators}
 We propose to estimate the latent vectors $\mathbf z_t$ from the SI navigators using an auto-encoder.  In this work, we applied a low pass filter with cut-off frequency of 2.8 Hz to the SI navigators to remove high-frequency oscillations. Similarly, an eighth-degree Chebyshev polynomial is fit to each navigator voxel and is subtracted from the signal to remove drifts. 

The auto-encoder involves an encoder that generates the latent vectors $\mathbf z_t = \mathcal E_{\varphi}(\mathbf y_t)$, are the navigator signals. The decoder reconstructs the navigator signals as $\mathbf y_t = \mathcal D_{\psi}(\mathbf z_t)$.  In this work, we restrict the dimension of the latent space to three, two corresponding to respiratory motion and one corresponding to cardiac motion. To encourage the disentangling of the latent vectors to respiratory and cardiac signals, we use the prior information on the range of cardiac and respiratory frequencies as in \cite{feng2016xd}.  We solve for the auto-encoder parameters from the navigator signals of each subject as 
\begin{equation}\label{autoencoder}
    \{\varphi^*,\psi^*\}=\arg\min_{\varphi,\psi}\left\| \mathcal{F}\left\{D_{\psi}\left(\underbrace{\mathcal E_{\varphi}(\mathbf Y)}_{\mathbf Z}\right) - \mathbf Y\right\}\right\|_{l=1} + \lambda~\left\|\mathbf Z\bigotimes\mathcal{B}\right\|_2^2
\end{equation}
Here, $\mathbf Z \in \mathbb R^{3\times T}$ and $\mathbf Y$ are matrices whose columns are the latent vectors and the navigator signals at different time points.  $\mathcal{F}$ is the Fourier transformation in the time domain.  $\bigotimes$ denotes the convolution of the latent vectors with band-stop filters with appropriate stop bands.  In particular, the stopband of the respiratory latent vectors was chosen to be $0.05-0.7$ Hz, while the stopband was chosen as the complement of the respiratory bandstop filter Hz for the cardiac latent vectors.  We observe that the median-seeking $\ell_1$ loss in the Fourier domain is able to offer improved performance compared to the standard $\ell_2$ loss used in conventional auto-encoders.

\subsection{Motion compensated image recovery}

Once the auto-encoder parameters $\varphi,\psi$ described in \eqref{autoencoder} are estimated from the navigator signals of the subject, we derive the latent vectors as $\mathbf Z = \mathcal E_{\varphi^*}(\mathbf Y)$. Using the latent vectors, we pose the joint recovery of the static image template $\boldsymbol \eta(\mathbf r)$ and the deformation maps as 
\begin{equation}\label{time}
    \{\boldsymbol \eta^*,\theta^*\} = \arg \min_{\boldsymbol \eta,\theta} \sum_{t=1}^T \|\mathcal A_t\left(\rho(\mathbf r,t)\right)-\mathbf b_t\|^2~~\mbox{where}~~ \rho(\mathbf r,t) = \mathcal I \left(\boldsymbol \eta,\mathcal G_{\theta}[\mathbf z_t] \right)
\end{equation}
The above optimization scheme can be solved using stochastic gradient optimization. Following optimization, one can generate real-time images shown in Fig. \ref{volunteer1} and Fig. \ref{volunteer2} as $\mathcal I \left(\boldsymbol \eta^*,\mathcal G_{\theta^*}[\mathbf z_t] \right)$.

\begin{figure}\label{overview}\centering
\includegraphics[width=0.7\textwidth]{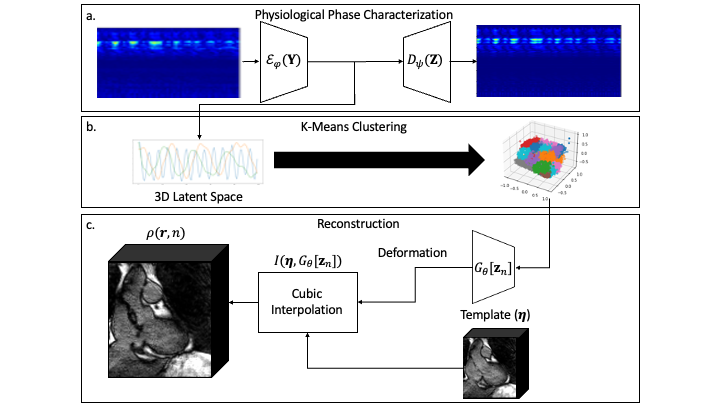}
\caption{Overview of the proposed reconstruction algorithm. In the first step shown in (a), we estimate the latent variables that capture the motion in the data using a constrained auto-encoder, as described in Fig. \ref{autoencoder}. The auto-encoder minimizes a cost function, which is the sum of an $\ell_1$ data consistency term and a prior involving cardiac and frequency ranges. To reduce the computational complexity of the image reconstruction, we cluster the latent space using k-means algorithm as shown in (b). The cluster centers are fed in as inputs to the CNN denoted by $\mathcal G_{\theta}$, which outputs the deformation maps $\mathcal G_{\theta}[\mathbf z_n]$.
We jointly optimize for both the template $\mathbf \eta$ and parameters $\theta$ of the generator.}
\end{figure}

The optimization scheme described in \eqref{time} requires $T$ non-uniform fast Fourier transform steps per epoch. When the data is recovered with a high temporal resolution, this approach translates to a high computational complexity. To reduce computational complexity, we introduce a clustering scheme. In particular, we use k-means clustering to group the data to $N<<T$ clusters. This approach allows us to pool the k-space data from multiple time points, all with similar latent codes. 
\begin{equation}\label{clustered}
    \{\boldsymbol\eta^*,\theta^*\} = \arg \min_{\boldsymbol\eta,\theta} \sum_{n=1}^N \|\mathcal A_n\left(\rho(\mathbf r,n)\right)-\mathbf b_n\|^2~~\mbox{where}~~ \rho(\mathbf r,n) = \mathcal I \left(\boldsymbol\eta,\mathcal G_{\theta}[\mathbf z_n] \right)
\end{equation}
The above approach is very similar to \eqref{time}. Here, $\mathbf b_n$, ${\mathcal A}_n$, and $\mathbf z_n$ are the grouped k-space data, the corresponding forward operator, and the centroid of the cluster, respectively. Note that once $N\rightarrow T$, both approaches become equivalent. In this work, we used $N=30$.  
Once the training is done, one can still generate the real-time images as $\mathcal I \left(\boldsymbol\eta,\mathcal G_{\theta}[\mathbf z_t] \right)$.  
 
 \subsection{Motion resolved 5D reconstruction for comparison}
\label{more}
We compare the proposed approach against a compressed sensing 5D reconstruction. In particular, we used the SI navigators to bin the data into 16 bins, consisting of four cardiac and four respiratory phases as described. We use a total variation regularization similar to \cite{feng2016xd} to constrain the reconstructions. We determined the regularization parameter manually to obtain the best reconstructions. 

We note that the dataset with 6.15 minute acquisition is a highly undersampled setting. In addition, because this dataset was not acquired with intermittent fat saturation pulses, it suffers from streaking artifacts that corrupt the reconstructions.  

\section{Results}
We show the results from the two normal volunteers in Fig. \ref{volunteer1} and 4, respectively. The images correspond to 2-D slices extracted from the 3D volume, corresponding to different cardiac and respiratory phases. We also show the time profile of the real-time reconstructions $\rho(\mathbf r,t) = \mathcal I \left(\boldsymbol\eta,\mathcal G_{\theta}[\mathbf z_t] \right)$ along the red line shown in the top row. We note that the approach can capture the cardiac and respiratory motion in the data. The different phase images shown in the figure were extracted manually from the real-time movies.
\newpage
\begin{figure}[h!]\centering\label{latentVector}
\centering\includegraphics[width=0.9\textwidth]{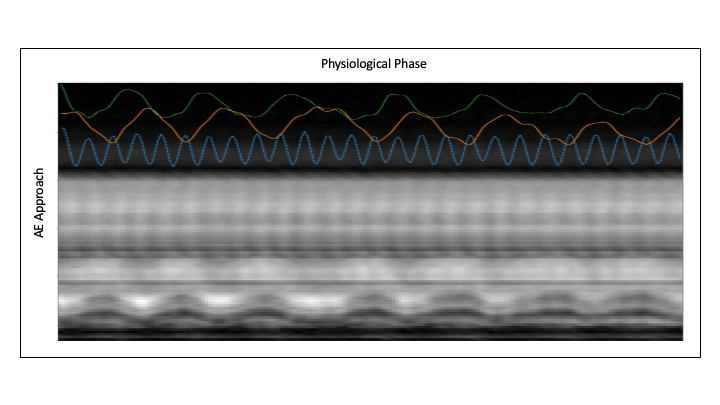}
\caption{Latent vectors estimated from the SI navigators (bottom curves)\cite{DiSopra2019}. We note that the orange and the green curves estimated using the auto-encoder roughly follow the respiratory motion, while the blue curves capture the cardiac motion.}
\end{figure}

\begin{figure}[h!]\centering\label{volunteer1}
\includegraphics[width=0.9\textwidth]{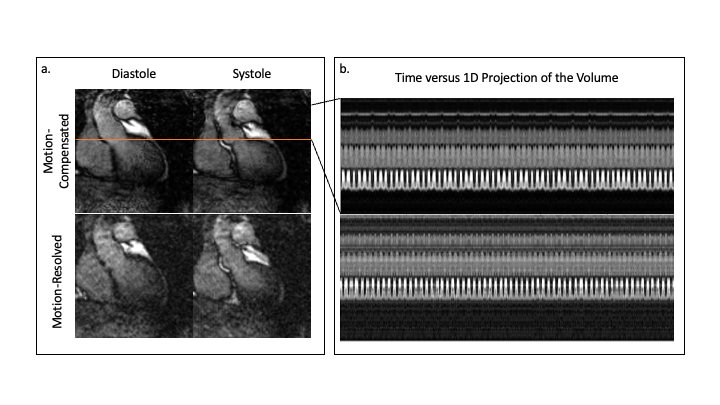}
\caption{Results from the first subject.  (a) The top row shows a 2-D slice of the reconstructed 3D volume at diastole and systole, obtained using the proposed motion compensated approach. The bottom row shows the motion-resolved compressed sensing recovery of the same data. (b) shows the 1D projection versus time profile of the reconstructed datasets using the motion compensated (top) and motion resolved (bottom) approaches.}
\end{figure}

\begin{figure}\centering\label{volunteer2}
\includegraphics[width=0.9\textwidth]{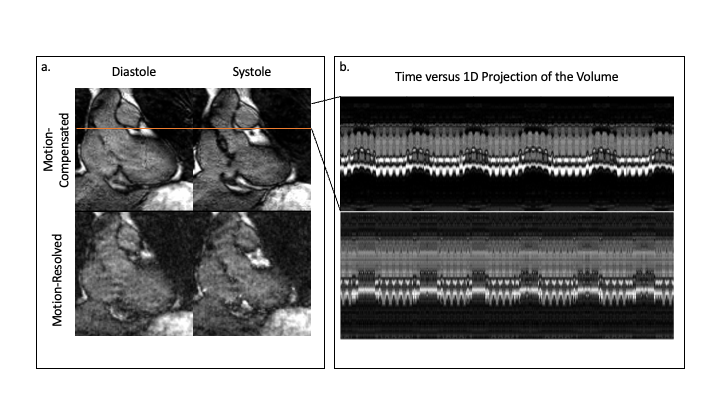}
\caption{Results from the second subject.  (a) The top row shows a 2-D slice of the reconstructed 3D volume at diastole and systole, obtained using the proposed motion compensated approach. The bottom row shows the motion-resolved compressed sensing recovery of the same data. (b) shows the 1D projection versus time profile of the reconstructed datasets using the motion compensated (top) and motion resolved (bottom) approaches.}
\end{figure}

\section{Discussion}
The comparisons in Fig. \ref{volunteer1} and 4 show that the proposed approach is able to offer improved reconstructions, where the cardiac phases are well-resolved. We note that the motion resolved reconstruction of the different phases have different image quality, depending on the number of spokes in the specific phases. By contrast, the proposed motion compensated reconstructions are able to combine the data from different motion states; the improved data efficiency translates to reconstructions with reduced streaking artifacts. Additionally, the auto-encoder accurately characterized the SI navigator and disentangled the cardiac and respiratory latent vectors Fig. 2.

We note that the comparison in this work is preliminary. The main focus of this work is to introduce the proposed motion-compensated reconstruction algorithm and the auto-encoder approach to estimate the latent vectors and to demonstrate its utility in 5D MRI. In our future work, we will focus on rigorous studies, including comparisons with 2D CINE acquisitions. 

\newpage
\bibliographystyle{splncs04}
\bibliography{refs}

\begin{thebibliography}{10}
\providecommand{\url}[1]{\texttt{#1}}
\providecommand{\urlprefix}{URL }
\providecommand{\doi}[1]{https://doi.org/#1}

\bibitem{Bustin2020}
Bustin, A., Fuin, N., Botnar, R.M., Prieto, C.: From compressed-sensing to
  artificial intelligence-based cardiac mri reconstruction. Frontiers in
  Cardiovascular Medicine  \textbf{7}, ~17 (2 2020).
  \doi{10.3389/FCVM.2020.00017/BIBTEX}

\bibitem{feng2016xd}
Feng, L., Axel, L., Chandarana, H., Block, K.T., Sodickson, D.K., Otazo, R.:
  Xd-grasp: golden-angle radial mri with reconstruction of extra motion-state
  dimensions using compressed sensing. Magnetic resonance in medicine
  \textbf{75}(2),  775--788 (2016)

\bibitem{Feng2017}
Feng, L., Coppo, S., Piccini, D., Yerly, J., Lim, R.P., Masci, P.G., Stuber,
  M., Sodickson, D.K., Otazo, R.: 5d whole-heart sparse {MRI}. Magnetic
  Resonance in Medicine  \textbf{79}(2),  826--838 (May 2017).
  \doi{10.1002/mrm.26745}, \url{https://doi.org/10.1002/mrm.26745}

\bibitem{Kim2021}
Kim, D., Cauley, S.F., Nayak, K.S., Leahy, R.M., Haldar, J.P.: Region-optimized
  virtual ({ROVir}) coils: Localization and/or suppression of spatial regions
  using sensor-domain beamforming. Magnetic Resonance in Medicine
  \textbf{86}(1),  197--212 (Feb 2021). \doi{10.1002/mrm.28706},
  \url{https://doi.org/10.1002/mrm.28706}

\bibitem{Mohsin2019}
Mohsin, Y.Q., Poddar, S., Jacob, M.: Free-breathing and ungated cardiac {MRI}
  using iterative {SToRM} (i-{SToRM}). {IEEE} Transactions on Medical Imaging
  \textbf{38}(10),  2303--2313 (Oct 2019). \doi{10.1109/tmi.2019.2908140},
  \url{https://doi.org/10.1109/tmi.2019.2908140}

\bibitem{Oscanoa2023}
Oscanoa, J.A., Middione, M.J., Alkan, C., Yurt, M., Loecher, M., Vasanawala,
  S.S., Ennis, D.B.: Deep learning-based reconstruction for cardiac {MRI}: A
  review. Bioengineering  \textbf{10}(3), ~334 (Mar 2023).
  \doi{10.3390/bioengineering10030334},
  \url{https://doi.org/10.3390/bioengineering10030334}

\bibitem{Piccini2011}
Piccini, D., Littmann, A., Nielles-Vallespin, S., Zenge, M.O.: Spiral
  phyllotaxis: The natural way to construct a 3d radial trajectory in {MRI}.
  Magnetic Resonance in Medicine  \textbf{66}(4),  1049--1056 (Apr 2011).
  \doi{10.1002/mrm.22898}, \url{https://doi.org/10.1002/mrm.22898}

\bibitem{Roy2022}
Roy, C.W., Sopra, L.D., Whitehead, K.K., Piccini, D., Yerly, J., Heerfordt, J.,
  Ghosh, R.M., Fogel, M.A., Stuber, M.: Free-running cardiac and respiratory
  motion-resolved 5d whole-heart coronary cardiovascular magnetic resonance
  angiography in pediatric cardiac patients using ferumoxytol. Journal of
  Cardiovascular Magnetic Resonance  \textbf{24}(1) (Jun 2022).
  \doi{10.1186/s12968-022-00871-3},
  \url{https://doi.org/10.1186/s12968-022-00871-3}

\bibitem{Rusho2022}
Rusho, R.Z., Zou, Q., Alam, W., Erattakulangara, S., Jacob, M., Lingala, S.G.:
  Accelerated pseudo 3d dynamic speech {MR} imaging at~3t using unsupervised
  deep variational manifold learning. In: Lecture Notes in Computer Science,
  pp. 697--706. Springer Nature Switzerland (2022).
  \doi{10.1007/978-3-031-16446-0\_66},
  \url{https://doi.org/10.1007/978-3-031-16446-0\_66}

\bibitem{Schlemper2018}
Schlemper, J., Caballero, J., Hajnal, J.V., Price, A.N., Rueckert, D.: A deep
  cascade of convolutional neural networks for dynamic mr image reconstruction.
  IEEE transactions on medical imaging  \textbf{37},  491--503 (2 2018).
  \doi{10.1109/TMI.2017.2760978},
  \url{https://pubmed.ncbi.nlm.nih.gov/29035212/}

\bibitem{DiSopra2019}
Sopra, L.D., Piccini, D., Coppo, S., Stuber, M., Yerly, J.: An automated
  approach to fully self-gated free-running cardiac and respiratory
  motion-resolved 5d whole-heart {MRI}. Magnetic Resonance in Medicine
  \textbf{82}(6),  2118--2132 (Jul 2019). \doi{10.1002/mrm.27898},
  \url{https://doi.org/10.1002/mrm.27898}

\bibitem{Uecker2008}
Uecker, M., Hohage, T., Block, K.T., Frahm, J.: Image reconstruction by
  regularized nonlinear inversion-joint estimation of coil sensitivities and
  image content. Magnetic Resonance in Medicine  \textbf{60}(3),  674--682 (Sep
  2008). \doi{10.1002/mrm.21691}, \url{https://doi.org/10.1002/mrm.21691}

\bibitem{Usman2013}
Usman, M., Atkinson, D., Odille, F., Kolbitsch, C., Vaillant, G., Schaeffter,
  T., Batchelor, P.G., Prieto, C.: Motion corrected compressed sensing for
  free-breathing dynamic cardiac mri. Magnetic Resonance in Medicine
  \textbf{70},  504--516 (8 2013). \doi{10.1002/MRM.24463}

\bibitem{Ying2007}
Ying, L., Sheng, J.: Joint image reconstruction and sensitivity estimation in
  {SENSE} ({JSENSE}). Magnetic Resonance in Medicine  \textbf{57}(6),
  1196--1202 (2007). \doi{10.1002/mrm.21245},
  \url{https://doi.org/10.1002/mrm.21245}

\bibitem{Zou2021}
Zou, Q., Torres, L.A., Fain, S.B., Higano, N.S., Bates, A.J., Jacob, M.:
  Dynamic imaging using motion-compensated smoothness regularization on
  manifolds (moco-storm). Physics in Medicine and Biology  \textbf{67} (12
  2021). \doi{10.1088/1361-6560/ac79fc}, \url{http://arxiv.org/abs/2112.03380
  http://dx.doi.org/10.1088/1361-6560/ac79fc}

\end{thebibliography}


%\end{thebibliography}
\end{document}